\documentclass{article}

\usepackage[utf8]{inputenc} % allow utf-8 input
\usepackage[T1]{fontenc}    % use 8-bit T1 fonts
\usepackage{hyperref}       % hyperlinks
\usepackage{url}            % simple URL typesetting
\usepackage{booktabs}       % professional-quality tables
\usepackage{amsfonts}       % blackboard math symbols
\usepackage{nicefrac}       % compact symbols for 1/2, etc.
\usepackage{microtype}      % microtypography
\usepackage{xcolor}         % colors

\usepackage{tabularx}
\usepackage{enumitem}
\usepackage{amsmath}
\usepackage{amsthm}
\theoremstyle{definition}
\newtheorem{definition}{Definition}[section]
\theoremstyle{remark}

\usepackage{graphicx} 
\usepackage{float} 
\usepackage{subfigure}
\usepackage{caption}
\usepackage{tabu}
\usepackage{mathstyle}

\PassOptionsToPackage{numbers,sort}{natbib}
\title{A Game-Theoretic Framework for AI Governance}
\author{Na Zhang, Kun Yue, Chao Fang \thanks{Corresponding author.}}
\date{November 2022}

\begin{document}

\maketitle

\begin{abstract}
    %Summarize motivations, problem to solve, model, results and core contributions of this work.
    As a transformative general-purpose technology, AI has empowered various industries and will continue to shape our lives through ubiquitous applications. Despite the enormous benefits from wide-spread AI deployment, it is crucial to address associated downside risks and therefore ensure AI advances are safe, fair, responsible, and aligned with human values. To do so, we need to establish effective AI governance. In this work, we show that the strategic interaction between the regulatory agencies and AI firms has an intrinsic structure reminiscent of a Stackelberg game, which motivates us to propose a game-theoretic modeling framework for AI governance. 
    %we leverage game theory to model the strategic interactions between AI corporations and governance agencies. 
    In particular, we formulate such interaction as a Stackelberg game composed of a leader and a follower, which captures the underlying game structure compared to its simultaneous play counterparts. Furthermore, the choice of the leader naturally gives rise to two settings. And we demonstrate that our proposed model can serves as a unified AI governance framework from two aspects: firstly we can map one setting to the AI governance of civil domains and the other to the safety-critical and military domains, secondly, the two settings of governance could be chosen contingent on the capability of the intelligent systems. To the best of our knowledge, this work is the first to use game theory for analyzing and structuring AI governance. We also discuss promising directions and hope this can help stimulate research interest in this interdisciplinary area. On a high, we hope this work would contribute to develop a new paradigm for technology policy: the quantitative and AI-driven methods for the technology policy field, which holds significant promise for overcoming many shortcomings of existing qualitative approaches. 

    %Lastly, we instantiate the Stackelberg governance framework by conducting a case study of autonomous driving.  

    %Game theory provides a systematic framework for modeling strategic interactions between key players across various domains. 
    %Simultaneous play vs. a hierarchical game with a play order.
   % capture the underlying game structure. 
    %stem from a distinct order of play between agents. 
    
\end{abstract}

\section{Introduction}
\label{sec:introduction}
Establishing an optimal governance framework of AI is very challenging due to many reasons. 
Besides the common challenges confronted by the technology policy field, AI governance poses a unique set of unprecedented problems. As AI is not a single kind of technology but a portfolio of diverse technologies, no silver bullet exists to solve the governance of all sub-sectors within the AI industry. Moreover, AI technologies are evolving quickly, which makes it challenging to analyze and track the effect of certain governance policies. Also, it indicates that the micro-level regulations should be adaptive, although the overarching principles should stay consistent. Besides, the governance agencies cannot manage and guide something they do not understand well. As AI becomes more complicated, the public policy makers struggle to catch up with state-of-the-art methods at the frontier. Last but not least, to tackle the governance of AI, we have to confront some fundamental ethical dilemmas that have been intensely debated. 

A common theme underlying these challenges is to understand the complicated interactions between the regulatory departments and the AI firms/institutions. In fact, the activities of the two parties are closely intertwined and the influence are usually bidirectional. It indicates that we must consider the strategic interactions of the two sides systematically in order to design effective AI governance policy and formulate a governing framework. Towards this end, this work takes a step by proposing a unified governance framework leveraging game theory. 

For a multi-agent system with self-interested agents, game theory is applied to analyze the optimal strategy for each agent to play. A large portion of the game theory literature focuses on simultaneous play, where all agents take actions at the same time. The standard solution concept for non-cooperative simultaneous play games is Nash Equilibrium \cite{nash1950}. However, for many real-world problems not only do the players assume asymmetric roles, but also take actions in an prescribed order. For such scenarios, hierarchical games that imposes a play order on the players are a better model. 
In fact, the design of a Stackelberg game enables it to capture such asymmetric structure of the play \cite{stakelberg2010}. 
In its simplest form, a Stackelberg game has two players, a leader and a follower: the leader is designated to act before the follower with the anticipation that the latter will play a best response. And the leader can uses this information to its advantage in figuring out her own optimal policy. In this light, the leader's strategy is 'the best response to the best response'. 

The focus of this work is to understand the interactions between the AI corporations and the regulatory agencies through the lens of game theory, which would shed light on formulating optimal governance of AI. %by accounting for the complicated interactions between the two sides. 
From the standpoint of game theory, the two players and their objectives are: (1) AI corporations/institutions aims to maximize return from commercialization of AI technologies; (2) AI regulatory ministries/departments seek to mitigate the potential downside risks due to AI research, development and adoption and meanwhile motivate AI innovation. The strategic interactions can be modeled as a game as the payoffs of each player depends on the parameters of both sides while each can only choose her own parameters. Furthermore, as the objective functions of the two are typically different (not simply opposite), this makes the game general-sum. 

In the context of AI governance, another important feature is that the players are asymmetric. So the symmetric solution concept Nash Equilibrium may not be desired for such situations. This observation again motivates us to resort to the Stackelberg game described above. In a nutshell, the hierarchical intersection between the governance agencies and AI corporations naturally lends itself to the Stackelberg game model. Given this abstraction, we cast the interaction of this pair as a Stakelberg game
%which is asymmetric game with a leader that plays first and a follower makes the decision after the former. [calculate an optimal decision by accounting for the anticipated best response of the follower.]

The asymmetric nature of Stackelberg games gives rise to two settings depending on the choice of the leader. By design, the player designated as the leader would act before players assigned as followers. Furthermore, the choice of the leader naturally gives rise to two settings; and we will demonstrate that our proposed model can serves as a unified AI governance framework as we can map one setting to the civil domains and the other to the safety-critical and military domains. 
%We will show that our proposal A unified governance framework for civil and military AI applications: (1) for civil domains, AI corporation/institutions as the leader and AI governance ministry/departments as the follower; and (2) for military domains, switch the roles of the two. 
Such an overarching framework help simplify the analysis from the lens of abstraction and unify multiple insights based on a principled foundation. Notably, our method deviates from the traditional approach for technology regulation. 

%Key technical steps: first, formulate the problem as a Stackelberg game; second, carefully constructs objective functions of the leader and the follower respectively; third, choose appropriate optimization approaches to solve the game; and lastly, characterize equilibrium properties of the game. 

\subsection{Main contributions}
The core contributions of this work are three-fold: (1) we view AI governance through the lens of game theory and cast the hierarchical interaction between AI firms and regulatory agencies as a Stackelberg game; (2) based this abstraction, we propose an overarching game-theoretic framework for AI governance, which unifies various existing insights and provides a principle model for understanding and analyzing complicated AI regulatory dynamics; furthermore, the Stackelberg equilibrium enables automatic balance between effectiveness and safety of AI innovation and application; (3) instantiate the theoretic framework by showcasing two special categories, incentive games and Stackelberg MDP, under the general framework, which clearly demonstrates the generality and flexibility thereof. \textbf{On a high level, we hope this work would contribute to initiate a new paradigm for technology policy research: the quantitative and AI-driven methods for technology policy, which can serve as a complementary approach to existing methods.} 

\subsection{Related Work}
\paragraph{AI Governance Framework}
We summarize most relevant recent research on AI governance frameworks, mostly proposed by the public policy community. 
\cite{layered_model2017} proposes a layered model for AI governance, which is composed of the technical layer, the ethical layer, and social \& legal layer. Another work
\cite{ai_governance_agenda2018} argues to divide the AI governance research into three clusters: the technical landscape, AI politics, and ideal governance of AI. Both frameworks recognize the foundational role of the technical part in the overarching framework. 
Besides surging academic attention, this area also attracts intense interest from the industry. For example, IBM explored AI governance framework and provides service
to help enterprises to use AI in a responsible and governed manner \cite{ibm_governance2022}; and highlights the proper AI governance would bring the organizations considerable benefits \cite{ibm_mature_governance2022}. 
\cite{integrated_governance_framework2020}
observes that government and public administration lags behind the fast development of AI and therefore fails to provide adequate governance in order prevent associated risks. To address this, \cite{integrated_governance_framework2020} develops an integrated framework that compiles crucial aspects of AI governance. 

%The dark sides of artificial intelligence: An integrated AI governance framework for public administration, 2020. 
%Main challenges of AI governance: 

\paragraph{General-Sum Stackelberg Game}
So far, Stackelberg game has found wide applications in various problems including modeling competing firms and first-mover advantage, allocating resources, improving public policy-making, and designing reinforcement learning algorithms, just to name a few. In particular, the Stackelberg game model has played an important role in security at infrastructure such as airports, transportation, to model the interactions between the security force and attackers and compute an optimal security force allocation strategy \cite{optimal_strategy_commit2006}\cite{security_game2011}. In particular, the IRIS program deployed by the US Federal Air Marshals is an well-known and successful example \cite{security_allocation2009}. 

In recent years, we see significant research interest in Stackelberg games partly due to these successful applications in modeling and analyzing various real-world problems. The majority of existing work focuses on zero-sum Stackelberg game [add references, while its general-sum counterpart is more relevant for many scenarios in practice. In zero-sum games, \cite{interchange2011} shows that Stackelberg and Nash equilibria are interchangeable, but the two typically diverge under the general-sum setting. Moreover, structural properties that lead to efficient computation of Nash equilibrium is not sufficient for the case of Stackelberg equilibrium, which shows the latter is more challenging to compute compared to the former \cite{interchange2011}. 

A line of recent work explores Stackelberg equilibria in the general-sum setting. 
\cite{first_order_stackelberg2019} investigate the properties of local convergence of the first-order methods for general-sum Stackelberg games; while \cite{bilevel_global_first_order2020} also focuses on the first-order methods, it studies the global convergence thereof for solving bi-level optimization problems. In a more challenging setting, where the reward functions and/or transition kernel are unknown, another line of recent research resorts to reinforcement learning methods for computing Stackelberg equilibria \cite{rl_general_sum2021} \cite{ai_economist2021}. 
%\cite{rl_general_sum2021} RL methods, general-sum, assume reward functions known  and transition model unknown and focus on the latter; 
%\cite{ai_economist2021} also RL methods; 
Among these, \cite{large_general_sum2019} explores sample-efficient algorithms for finding Stackelberg equilibrium of large general-sum games where the leader and/or the follower action set is exponential in the natural representation of the problem. 

%\cite{oracle_auction2016} when a best-response oracle is available; 

\paragraph{Multi-Agent Reinforcement Learning}
Multi-agent reinforcement learning (MARL) studies the sequential decision-making problem involves multiple interacting learning agents, which is an extension of the single-agent RL. It is typically modeled as a Markov game \cite{markov_game1994}, during each step of which all players take actions simultaneously. Furthermore, the relationship among agents could be either collaborative or competitive, contingent on the structure of their reward functions. So far, single-agent RL has achieved prominent success and been widely applied in solving various tasks \cite{mnih_control_nature2015} \cite{control_drl2015}   \cite{silver_go_nature2016}. However, many real-world scenarios, such as autonomous driving and multi-robot control, naturally comprise more than one decision-makers that collectively learn, interact, cooperate and/or compete in a shared environment, which could be modeled as a multi-agent RL problem. 

Compared to its single-agent counterpart, a number of exclusive challenges arise in the multi-agent RL setting: non-stationarity, learning communication, coordination, credit-assignment, scalability, and partial observation \cite{multiagent_rl_survey2022}. In particular, non-stable learning as well as co-adaptation of the interacting agents makes the sequential decision-making very challenging. In addition, the sample complexity of centralized multi-agent RL algorithms scales exponentially with the number of agents. To address this challenge, a line of recent research \cite{multi_rl_v2021} \cite{v_decentralized2021} proposes a decentralized structure, V-learning, for finding equilibrium strategy. 
%A line research seeks to generalize well-performing single-agent RL algorithms to the multi-agent setting [add references]. 

To date, most of recent literature on multi-agent RL focuses on finding Nash equilibrium in the Markov game without taking the asymmetric structure of players into account. Furthermore, for recent work that do explore Stackelberg equilibrium, the majority works on the zero-sum case. 

\paragraph{Automated Mechanism Design}
Mechanism design studies how to create preference aggregation rules that can produce desirable outcomes in spite of self-interest agents, while automated mechanism design aims to generate a general mechanism for the problem of interest in an automatic manner, usually based on constrained optimization methods \cite{automated_mechanism2007}. The latter attracts wide interest for researchers in economic and public policy fields for a number of factors. First and foremost, many real-world policy optimization poses a mechanism design problem. In addition, the computational methods for automated mechanism design address some of the aforementioned challenges confronted by mechanism design, such as lack of experimental data, the Lucas critique, etc. Recent work further proposes methods for solving multistage mechanism design automatically, which can be formulated as optimization problems \cite{automated_mechanism2007}. 

Compared to the aforementioned Stackelberg games, note that the designer actually has an implicit leadership position in the mechanism design setting.

\paragraph{Organization} Lastly, we briefly summarize the arrangement of the rest of this paper: section 2 provides brief background on AI governance and Stackelberg game; based on this, section 3 introduces our Stackelberg framework for AI governance; we then instantiate the general framework and discuss two special cases fit the AI governance problem naturally; we end by summarizing the main message and pointing out a number of directions for future work.

\section{Preliminary}
\label{sec:preliminary}
% proivde background, second part provides relevant game theory backgournd and formalism. 

\subsection{AI Governance} 

In general, AI governance refers to 'the ability to direct, manage and monitor the AI activities of and organization' \cite{ibm_governance2022}, the ultimate goal of which is to ensure AI is trustworthy and responsible and hence benefits humankind. AI governance has recently attracted interest from both governments and academia globally. 

Besides the common challenges confronted by technology policy field, AI governance poses many unique challenges. First of all, AI is not A kind of technology but a portfolio of diverse technologies, which implies that there is no silver bullet that fit all sub-sectors within AI. Secondly, AI technologies are evolving quickly as this field as a whole is still in its early stage. This indicates that micro-level regulation should be adaptive, although the overarching principles stay consistent. Moreover, we cannot manage and guide something we do not understand. As AI becomes more complicated, public policy makers might struggle to catch up with state-of-the-art methods at the AI frontier. Last but not least, to tackle AI governance, we have to confront some ethical dilemmas not seen before. 
Apparently, interdisciplinary experts are crucial to create new wisdom for this cross-field arena. On the other hand, most traditional tools, mostly qualitative and developed by the public policy researchers, are inadequate to effectively address aforementioned challenges. This motivates us to explore new methodologies for optimal governance of AI. 
% the lack of interdisciplinary experts that combines deep understanding of AI and technology management 
On a high level, we hope this work would contribute to initiate a new paradigm for technology policy research: the quantitative and AI-driven methods for technology policy, which can serve as a complementary approach to existing methods.

\subsection{Stackelberg Game}
Stackelberg games are hierarchical games with a leader-follower structure and a prescribed play order: the leader(s) plays before the follower and chooses a strategy that maximizes her rewards, acknowledging that the latter can observe the her action and play best response. In this light, the leader's strategy can be viewed as 'the best response to best response'. In essence, Stackelberg game is a generalization of min-max games, and it can technically be formulated as a bi-level optimization problem. In contrast to the case that each player optimizes her objective independently, the formulation of Stackelberg games takes the hierarchical structure into account, which gives rise to a number of appealing properties and makes it prevalent to many real-world scenarios summarized in section 1. 

We begin by highlighting a few aspects of the game design that are most relevant to our work. 
\begin{itemize}
    \item \textbf{zero-sum vs. general-sum} When the reward functions of the leader and follower are not opposite, general-sum is more prevalent compared to the simpler but more constraint zero-sum setting. 
    \item \textbf{normal form vs. extensive form} In normal form games, players choose a strategy with beliefs about how other players act, which can be represented by a table. Extensive form games allow each player to access a complete description of all previous strategies that other players have chosen so far and in what order. We focus on the normal form throughout this work. 
    \item \textbf{optimistic tie-breaking vs. pessimistic tie-breaking} Given an action of the leader, there could be more than one best response strategies of the follower, which constitute a best response set. Optimistic tie-breaking means the follower would choose an action from the best response set in favor of the leader, while pick one adversarial to the leader in the pessimistic tie-breaking. The AI governance frame in section 3 adopts the former. 
    \item \textbf{one leader and one follower vs. multiple leaders and/or multiple followers} In its simplest form, there is one leader and one follower. We work on this two-player case in this work and discuss potential extensions to multiple interacting followers when discussing future work. 
    
\end{itemize}
 
%Stackelberg game vs. Markov game; 
%zero-sum vs. general-sum; 
%one leader one follower vs. one leader multiple followers; 
%global optimal solution vs. local optimal solution; 
%optimistic tie-breaking vs. pessimistic tie-breaking; 
%single-step game vs. multi-step game; 

So far, Stackelberg game has found wide applications in various problems including modeling competing firms and first-mover advantage, allocating resources, improving public policy-making, and designing reinforcement learning algorithms, just to name a few. In this section, we provide a formal representation of Stackelberg game.

\paragraph{Formulation}
Here, we provide the notations for Stackelberg game. Use $\theta$ to denote the variables that the leader can control, while $\omega$ refers to parameters that the follower can set. Suppose the objective functions of the leader and follower are $J(\theta, \omega)$ and $L(\theta, \omega)$ respectively; both aim to, without loss of generosity, minimize their objective functions by making choices of $\theta$ and $\omega$ in a prescribed order. By design, the leader would move first, and then the follower would play best response; the leader knows this and would take it into consideration when figuring out her own optimal decision.

We firstly define the best response set of the follower, denoted as $BR(\theta)$,  and the optimal parameter $\omega^*$ under the optimistic tie-breaking by: 
\begin{equation}
    BR(\theta) := \{\omega | \omega \in \argmin_{\Tilde{\omega} \in \W}L(\theta, \Tilde{\omega}) \}, 
\end{equation}
\begin{equation}\label{br}
    \omega^*(\theta) := \{\omega \in BR(\theta) | J(\theta, \omega) \le J(\theta, \omega'), \forall \omega' \in BR(\theta) \}
    %\argmin_{\omega \in BR(\theta)} J(\theta, \omega). 
\end{equation}

Then, in Stackelberg games, the leader and follower solve the following optimization problems in order: 
\begin{equation}
    \min_{\theta \in \Theta }J(\theta, \omega^*(\theta)), 
\end{equation}
\begin{equation}
    \min_{\omega \in \W}L(\theta, \omega). 
\end{equation}

\paragraph{Equilibrium Characterization} 
We begin by presenting the equilibrium concept for simultaneous play games and then contrast it with its counterpart in Stackelberg games. For the former, Nash equilibrium(NE) is the standard solution concept, which specifies a profile of strategies that no player has the incentive to deviate from it. The seminal work by John Nash shows the existence of such equilibrium in finite games (there could be multiple Nash equilibria) \cite{nash1950}. In fact, most existing work focuses on such symmetric solution concept for zero-sum games. 

\begin{definition}[Nash Equilibrium (NE)]

    The joint strategy $(\theta^*, \omega^*)$ is a Nash equilibrium on $\Theta \times \W $, if $J(\theta^*, \omega^*) \le J(\theta, \omega^*), \forall \theta \in \Theta$, and $L(\theta^*, \omega^*) \le L(\theta^*, \omega), \forall \omega \in \W$. 
    
\end{definition}

Stackelberg equilibrium is the asymmetric counterpart of NE under the hierarchical games with an imposed play order. Accordingly, we provide the analogous equilibrium concepts in the hierarchical structure. 

\begin{definition}[Stackelberg Equilibrium (SE)]\label{se}

The strategy $\theta^*$ is a Stackelberg solution for the leader if $sup_{\omega \in \omega^*(\theta^*)} J(\theta^*, \omega) \le sup_{\omega \in \omega*(\theta)}J(\theta, \omega), \forall \theta \in \Theta$; if $\omega^*$ is the follower's best response strategy under optimistic tie-breaking defined in Eq.~\ref{br}, then the strategy profile $(\theta^*, \omega^*)$ is a Stackelberg equilibrium on $\Theta \times \W$. 

\end{definition}

For zero-sum games,  von Neumann's minimax theorem shows that Nash equilibrim coincides with Stackelberg equilibrium, but the two diverge significantly in the general-sum setting \cite{interchange2011}. For Stackelberg games, \cite{stackelberg_nash2010} shows that the leader's utility in a strong Steckelberg equilibrium is at least as high as it in any Nash equilibrium. 
However, in many real-world cases, the players do not select their strategies in a simultaneous manner, but have to follow a specified order, which naturally fit to the Stackelberg model. 
%While Nash equilibrium is the same with Stackelberg equilibrium under the zero-sum finite game setting, the two could be significantly different in the general-sum case \cite{interchange2011}. 

When the global optimal strategy is infeasible, a weaker version of SE, local Stackelberg equilibrium, is defined on local regions of the players' action spaces and formalizes the concept of local optimal solution. 

\begin{definition}[Local Stackelberg Equilibrium (LSE)]\label{lse}

Let $\Theta_{1} \subset \Theta$ and $\W_{1} \subset \W$. The strategy $\theta^{*} \in \Theta_{1}$ is called a local Stackelberg equilibrium for the leader if $\forall \theta \in \Theta_{1}$, 
\begin{equation}
    \sup_{\omega \in R_{W_{1}}(\theta^*)} J(\theta^*, \omega) \le \sup_{\omega \in R_{W_1}(\theta)} J(\theta, \omega), 
\end{equation}
where $R_{W_1}(\theta):=\{x \in  W_1 |L(\theta, x) \le L(\theta, \omega), \forall \omega \in W_1\}$. $\forall \omega^* \in R_{W_1}(\theta^*)$, $(\theta^*, \omega^*)$ is a local Stackelberg equilibrium on $\Theta_1 \times \W_1$. 

\end{definition}

% Differential Stackelberg Equilibrium is for zero-sum game. 

\paragraph{Methods for Solving Stackelberg Equilibrium}
%[global optimal solution vs. local optimal solution; ]

Following this, we briefly discuss recent progress in computing Stackelberg equilibrium in general-sum games. Note that formally characterizing the existence of Stackelberg equilibrium is out of the scope of this work. We refer the readers for (Barsar and Olsder 1998, these equilibria can be characterized as the intersection points of the reaction curves of the players and shows that they exist on compact strategy spaces.)

There has been an extensive literature on learning equilibria in zero-sum Stackelberg games \cite{stakelbergdynamics2020}. 
%[Add related work.] 
By contrast, theoretical understanding for the general-sum setting remains vastly open, despite a number of recent work in this direction. When the game size is small, SEs can be computed efficiently; and a popular method is Multiple LPs proposed in \cite{optimal_strategy_commit2006}. However, for large general-sum Stackelberg games, existing approaches often fails to compute the equilibrium concepts efficiently. As highlighted earlier, Stackelberg and Nash equilibria typically diverge in the general-sum scenarios. %Can RL find Stackelberg/Nash equilibrium with myopic followers? 
\cite{rl_general_sum2021} uses RL to find SE with myopic followers (only consider immediate reward of an action and ignore future rewards) in both online and offline settings, which are optimistic and pessimistic variants of least squares value iteration respectively. For the case of linear function approximation, their algorithm can achieve sublinear regret and suboptimality in general-sum Markov games under the assumption of myopic followers. 

A concurrent work in this line \cite{sample_efficient2021} also leverages RL to address the challenge of learning SE given only bandit feedback without knowing the reward functions and the transition model. In this bandit-RL setting, they identify a gap between the true value of SE and its estimate under finite samples from an information-theoretical perspective. However, this work shows that it is possible to construct sample-efficient algorithms with value optimal up to the identified gap. 
In addition, \cite{ai_economist2021} frames computing optimal tax policy as a bi-level RL problem, and then combines machine-learning-driven simulation and RL methods to find SE solution. 

%Computing Stackelberg equilibrium of large general-sum games \cite{large_general_sum2019}: Computing equilibrium for general-sum Stakelberg games.  

%Learning Stackelberg equilibrium in general-sum games 
%\cite{sample_efficient2021}

\section{Stackelberg Framework for AI Governance}
\label{sec:framework}
As AI becomes a shaping power for our and future generations, AI research and development has becomes a top priority globally. While too much regulation might hinder AI innovation, the governance is crucial to ensure AI would be safe, trustworthy, and responsible. In this section, we show that the abstraction of AI governance problem that naturally lends itself to a Stackelberg game structure that couples 1) AI innovation, development and adoption and 2) AI regulatory. Then, building off this abstraction, we develop a game theoretical framework towards the optimal governance of AI. Towards this goal, the primary issue is to better understand and model the interactions between AI firms and regulatory agencies, and our primary contribution lies in casting it as a general-sum Stackelberg game. 

While most of the contemporary work focuses on simultaneous play games and corresponding Nash equilibrium, the structure of Stackelberg games apparently  makes it a much better fit in the context of AI governance for a number of reasons: 
\begin{itemize}
    \item The governance agencies and AI firms apparently assumes different roles and make their decisions at different time. These facts fit well with the interaction structure, asymmetric players and a specified play order, in Stackelberg games; 
    \item In practice, AI firms would choose the business strategy to maximize its profit, both of which match the best response strategies in Stackelberg games; 
   % \item The best response to best response for leader
    \item Stackelberg games with multiple interacting followers can be used to model AI firms in the market; 
    \item The choice of the leader in the Stackelberg games naturally gives rise to two different settings, which could be mapped to the governance of two general categories of AI sub-sectors. 
\end{itemize}

Together, the above aspects shows that the interaction between the regulatory agencies and AI firms has a intrinsic structure reminiscent of a Stackelberg game, which motivates us to propose a game-theoretic modeling framework for AI governance.

Again, while we adopt the one leader and one follower setting throughout for simplicity, note that the framework can be extended to incorporate the multiple followers case. 

\paragraph{Setting Up and Notations} Let $\pi$ denote the corporate strategy ranging from AI investment, research and development, to deployment, which can be chosen from a strategy space $\Pi$, continuous and/or discrete. We assume that the corporate players establish their AI strategies and business decisions based on the evaluation of AI performance along a suite of key dimensions integrated  in $\mu$, and the cost as well as constraints due to the regulation and standards set by the AI governance departments, denoted $c$. 

In this framework, players evaluate the AI technologies from six dimension aspects: performance, robustness, explainability, fairness, privacy, and security. (A future direction is to develop better metrics to measure AI.) The AI development strategy is a primary factor that determines the overall development level of AI, and therefore we can make $\mu$ a function of $\pi$. The cost and constraints imposed on AI firms by regulation agencies can be derived from the performance for AI captured by $\mu(\pi)$ and the regulations and standards $\omega$, and we denote it as $c(\mu, \omega)$. We emphasize that $\pi$ refers to the variables that the corporate can control, while $\omega$ are variables that the AI regulatory agencies can set. 

Equipped with this, we are ready to set up the objective functions for the AI corporations and governance agencies. The former seeks to maximize the return from AI innovation and application. Note that return here is defined in the broadest sense, covering not only financial benefits but also intangible assets like competitiveness, not only short-term return but also long-term value. 
Formally, the objective function can be modeled as follows: 

\begin{align}\label{return}
    J_{\pi, \omega} := J(\mu(\pi), c(\mu,\omega))
\end{align}

On the other hand, the goal of the regulation player is to minimize the downside risks due to AI development and deployment. To this end, the agency are motivated to minimize a loss function that depends on both the AI development level and regulations, 
formally defined as

\begin{align}\label{loss}
    L_{\pi, \omega} := L(\mu(\pi), \omega) 
\end{align}

Furthermore, we argue that there are no one-fit-all objective functions for both players, and such functions vary across domains. And we will choose autonomous driving as a case study in section 4. 

\paragraph{Stackelberg Game Formulation}
Our choice of the leader naturally gives rise to two forms of Stackelberg governance framework. Choosing AI corporations as the leader leads to the formulation as below: 

\begin{equation}
    \max_{\pi \in \Pi }J(\mu(\pi), c(\mu, \omega^{*}(\pi))) ~~ \text{s.t.}~~ \omega^*(\pi) \in \argmin_{\Tilde{\omega} \in \Omega}L(\mu(\pi),  \Tilde{\omega}), 
\end{equation}
\begin{equation}
    \min_{\omega \in \Omega}L(\mu(\pi), \omega)
\end{equation}

Alternatively, if we make the governance agency as the leader the AI corporations the follower, the setting up becomes the following nested optimization problem: 

\begin{equation}
    \min_{\omega \in \Omega }L(\mu(\pi^*),  \omega)  ~~ \text{s.t.}~~ \pi^*(\omega)=\argmax_{\Tilde{\pi} \in \Pi}J(\mu(\Tilde{\pi}), c(\mu(\Tilde{\pi},\omega)), 
\end{equation}

\begin{equation}
    \max_{\pi \in \Pi}J(\mu(\pi), c(\mu(\pi), w))
\end{equation}

The two formulations clearly shows that the Stackelberg games are asymmetric. \textbf{Importantly, we highlight the fact that the leader can benefit from the hierarchical play order compared to simultaneous play games. Therefore, which agent should be deemed as the leader is a critical decision contingent on the industry of interest.}
In fact, the above two forms typically lead to different equilibria and we argue that they are suitable for analyzing the governance of distinct subsectors of AI, as elaborated next.

\subsection{Choose Governance Settings Based on Domains}

\subsubsection{Civil Domains: AI Corporations as Leader}
The rise of AI began from the civil domains. We witness the power of 'the invisible hands' of the market in terms of promoting AI innovation and commercialization. From the viewpoint to maintain continuous creativity in AI, it is critical to provide enough space for firms in this area to allow for exploration and experiments. Moreover, the potential cost and soical impact of experiments in civil domains is relatively controllable. For instance, when the AI department at Alibaba or Amazon leverages machine learning algorithms for making recommendations on which products to buy for customers, the cost of a recommendation that fails to meet the customer's need is limited.  

On the other hand, a well-known fact is that it is very challenging for the governance ministry to foresee AI development direction and possible risks, which are essential to formulate optimal guidance thereof. 
Combining these factors, we argue to assign the role of leader in the Stackelberg game formulation to the AI corporations in the civil domains.

\subsubsection{Safety-Critical and Military Domains: AI Regulation Agency as Leader}
A number of issues distinguish the safety-critical and military industries from the civil domains when it comes the regulation of the AI technology research and application. This indicates we should reconsider the roles of the players for these sectors. 

%AI safety concerns. 
First of all, AI safety is a primary concern for both safety-critical and military domains. To apply AI in such industries like medical care, financial investment, as well as autonomous driving, allowing the firms to act without having certain governance in place could be dangerous and induce high cost for the society as a whole. All the more so for the military field. 

%a support but not an alternative; to empower not to replace; 
In particular, there are numerous scenarios involves complicated and high-stakes decision-making, however, even the state-of-the-art AI technologies may not be transparent and robust enough to replace human beings in such cases. Take nuclear weapon systems for example, a key advantage of machine learning lies in automatically collecting large-scale unstructured raw data, processing them efficiently, and mining the underlying relationships of among various factors, and then making informed predictions in an end-to-end manner. These attributes make machine learning a very attractive tool for many operations in the early-warning and ISR. However, due to the fundamental constraints of contemporary AI, it is not ready to replace human beings in the command and control section within the nuclear deterrence structure. 

Apparently, implementing a wrong decision could be extremely costly and/or the result might even be irreversible. Moreover, AI corporations and institutions do not fully internalize the entire cost of downside risks. For instance, AI firms can reap the short-term profits from AI application, they tend to downplay long-term social or systematic risks. Hence, to cope with the costly potential risks and failures, it is essential to that AI governance for such areas should be relatively more prudent. 

%establish appropriate standards for AI implementation at the first place. 
Taking these factors into account, we argue that it is desirable to deem the AI governance agency as the leader in the Stackelberg framework for both safety-critical and military domains. Thus, agencies can act first by setting regulation rules and standards before organizations implement AI or other relevant activities. 

\subsection{Choose Governance Settings Contingent on AI Capabilities}

The advent of large-scale AI models (a.k.a. foundation models), LLMs in particular, brings a big leap in performance and triggers a paradigm shift of the entire AI field. While being stunned by the capabilities of the new generation of AI systems, we have to confront multiple downside risks that are unprecedented. We argue that it is crucial to distinguish the governance of this dramatically more capable AI systems from that of the traditional machine learning techniques for a number of factors discussed in the succeeding content. 

To do so, the first step is to establish a threshold of AI capability. Given this, once the capability of the intelligent systems surpasses the designated threshold, a more prudent governance setting with the regulatory agency assumes the leader role in the Stackelberg game applies; otherwise, the other setting with the industry as the leader is applicable by default. Foreseeing the incoming powerful intelligent systems, we highlight a number of primary issues under the general governance framework in the former case below. 

\subsubsection{Governance of the New Generation of AI Systems}

While the rise of foundation models has made impressive achievements across a multitude of diverse downstream tasks, we should be aware of the other side of the coin: potential downside risks and structural changes to the society. In particular, we still do not have satisfying ways to contend with model misuse 
and to align models with desirable human values. Recognizing these, it is imperative to assign the leader role to the regulatory agencies when formulating governance polices for the new generation of AI systems, which are dramatically more powerful than before. 

Under the general Stackelberg game-theoretic framework, formulating the objective functions of the players is at the core, which is closely related to the goals, challenges of governance, and the properties, limitations and impacts of AI,  etc. In the AI governance setting with the regulatory agencies as the leader, we highlight some primary factors that the objective functions should seek to effectively characterize. 

\begin{itemize}

    \item The academic, industry and government agencies should work together to explore and establish the overarching principles of the AI governance framework. 
    \item Effective governance of the highly capable AI systems entails international cooperation. Specifically, it is helpful to establish a new international organization to coordinate the efforts towards effectively handling potential risks generated by this new generation of AI. 
    \item It is critical to manage and guide the integration of powerful AI systems into the society, in order to ensure the structural transition on a social level is smooth and controllable.  
    We already have a glimpse of incoming subversion from a number of areas hit by the wave of highly capable large-scale generative models. It is imperative to have a policy framework that can guide the society to navigate the approaching revolution. Among this, manage the pace of the transition, make the labor force be prepared for a totally different employment market. 
\end{itemize}

%Entrenching inequality, aggravating social fragmentation, exacerbating violence, country conflicts. 

Last but not least, it is worthy to note the discussed two criteria for choosing AI governance settings would naturally give rise four combinations.

\subsection{Computing Stackelberg Equilibria}
The appropriate method to compute the SEs depend on the assumptions or constraints imposed on the setting, such as whether the reward functions and transition kernel are known or not and whether it is possible to conduct interactions and generate new data. For our case, to find the LSE, it is possible to leverage a gradient-based learning algorithm to emulate the Stackelberg game structure. However, characterizing available methods for calculating SEs is outside the scope of this study and we leave it for future work.  

%Specifically, we outline a two-time scale gradient-based learning rules. 

%Known reward functions and transition model; 
%Unknown reward functions and transition kernel; 
%Able to conduct interactions and generate data;
%Limited opportunity to conduct experiments; 

\section{Instantiation of the General Framework}
\label{sec:instantiation}
In this section, we showcase two special categories under the general framework proposed in section 3, which clearly demonstrates the generality and flexibility thereof. 

\subsection{Incentive Games}
Incentive games can be viewed as a subcategory of Stackelberg games. Every action of the leader are composed of two parts: (1) an element of a set of actions that the leader can take, and (2) an incentive set designed to stimulate the follower to play certain actions. A primary property of incentive games is that the additional incentives that only benefit the follower can actually improve the leader's total utility in equilibrium. In practice, incentive games find applications in various domains. For example, 
\cite{incentive_energy2014} formulates the interaction between the building manager and the occupants as an incentive game and leverages the incentive design to encourage energy efficiency behavior of the latter. 
By parameterizing the players' utility functions and incentives, \cite{adaptive_incentive2018} employs control theoretic and optimization techniques to adaptively design incentives. Furthermore, \cite{large_general_sum2019} shows that there exist polynomial time algorithms for computing the SEs, 
under certain technical conditions. 

We see a natural fit between the model of incentive games and the case of AI governance. First and foremost, the regulatory agencies have interest and capability to design a combinatorial of incentives to guide and motivate the AI enterprises to make business decisions that meet socially beneficial goals and long-term visions, instead of solely based on short-term financial profits. Furthermore, as highlighted above, while these incentives are tailored to benefit the firms, well designed ones can ultiamtely lead to higher utility for the goverance agencies in SEs. In this light, incentive games provide a principled way to search for socially optimal governance policy. 

Formally, we denote the leader action $\pi := (a, \Vec{v})$, where the first part $a \in \A$ is an element from a set composed of all possible actions the leader can take and the other part $\Vec{v} \in [0, 1]^{|\A|}$ is a vector of incentives designed to motivate the follower to take certain actions. The design of the incentives is a crucial step. Given this, it is straightforward to update the objectives functions of the leader and follower in section 3 accordingly. 

\begin{align}
    J_{(a, \Vec{v}), \omega} := J(\mu(a, \Vec{v}), c(\mu,\omega))
\end{align}
\begin{align}
     L_{(a, \Vec{v}), \omega} := L(\mu(a, \Vec{v}), \omega) 
\end{align}

\subsection{Stackelberg MDP}
Given the policy established by the AI governance agencies, the firms typically would not implement a portfolio of business decisions all at once but gradually take a sequential of actions over time. This fact motivates us to extend the turn-based Stackelberg game in the general framework in order to model the sequential decision-making of the follower. Towards this end, we assume that each leader's action induces an episodic Markov Decision Process (MDP) of the follower. Specifically, in an episode of the game, the leader firstly commints to a strategy; the follower then enters a MDP after seeing the former's action, and observes a reward at every step of the MDP until it terminates, while the leader also observes her rewards at each step during this process. The objectives of both sides remains the same: finding policies that maximizes their accumulated rewards respectively. We name this setting Stackelberg MDP.  

Different from incentive games above, this time the leader's action space $\pi$
remains the same with that in the general framework, while the model of the follower is revised to represent the sequential decision-making thereof. Towards this end, we introduce a family of MDPs represented by a tuple of six elements, $(\Sset, \Aset_{f}, H, r_{l}, r_{f}, \P)$, where $\Sset$ is the state space, $\Aset_{f}$ the action space of the follower, $H$ the horizon of the follower MDP, $r_{l}$ and $r_{f}$ the reward functions of the leader and follower respectively, and $\P$ the transition kernel. The leader strategy $\pi$ induces a episodic MDP of the follower. The objectives of the two are find optimal strategies to maximize their expected cumulative (discounted) rewards respectively. 

%In challenging case of only receiving bandit feedback, \cite{sample_efficient2021} proposed an algorithm for computing the Stackelberg equilibrium. 

%\section{Quantitative Case Study: Autonomous Driving}
%\label{sec:case}
%\input{case.tex}

\section{Conclusion and Future Directions}
In this paper, we propose a game theoretic framework for (1) understanding the strategic interactions between the AI governance agencies and enterprises, and (2) designing optimal regulatory policies that ensures the development and deployment of AI is safe, responsible, and trustworthy. We wish this work would ignite interest for researchers from both AI and public policy communities, and therefore help open up a line of research at this promising interdisciplinary area. 

Furthermore, on a high level, we hope this work would contribute to develop a new paradigm for technology policy research: the quantitative and AI-driven methods for technology management. Most existing methods are qualitative, developed by researchers in the public policy community. The traditional quantitative methods typically make simplifications and assumptions for analytic tractability. As a result, the theoretical models fail to capture the real-world complexity. Besides, the theoretical models are hard to validate in practice. Recall that a primary challenge is lack of real data. However, even when real data is available, conclusions based on such historical data suffers from the Lucas critique. 

In this light, it is crucial to create new wisdom for this new interdisciplinary field. As the traditional tools are inadequate to effectively address aforementioned challenges in AI governance, this motivates us to explore the AI-driven methods for the optimal governance of AI, which would serve to provide a complementary approach to the existing qualitative work, not replacing them.

We end by discussing promising directions to explore in the future. We plan to instantiate our governance framework and leverage machine learning methods to train an AI governor to help learn dynamic and adaptive regulatory rules for AI. In particular, we would utilize deep reinforcement learning to search for optimal AI regulations. The AI governor is trained to discover regulatory policy that can optimally balance AI innovation and development with desirable social welfare including safety, ethics, fairness, etc. Moreover, as there is no one-fit-all solution for all sub-sectors within the AI industry. It is important apply the framework to conduct in depth case study of key areas of AI, such as autonomous driving. 

%AI-driven simulation: carefully construct the objective functions for AI corporations player and regulatory agencies player separately. Based on this, we can solve the game as a bi-level optimization problem. design, analyze and compare the equilibrium properties and implications of different governance policies.

%Lastly, while this work focuses on outlining the theoretical framework, which aims to lay out a foundation for future work in this interdisciplinary area, we also discuss promising directions and hope this can stimulate research interest thereof. 

\bibliography{game_theory}

\end{document}